\pdfoutput=1

\documentclass[sigconf]{acmart}

\usepackage{graphicx}
\usepackage{hyperref}
\usepackage{color}
\usepackage{todonotes}

\usepackage{amsmath}
\usepackage{algorithm}
\usepackage[inline]{enumitem}

\AtBeginDocument{%
  }

\setcopyright{acmlicensed}
\copyrightyear{2024}
\acmYear{2024}
\acmDOI{XXXXXXX.XXXXXXX}

\acmConference[JCDL '24]{The ACM/IEEE-CS Joint Conference on Digital Libraries}{December 16--20,
  2024}{Hong Kong, China}
\acmISBN{978-1-4503-XXXX-X/18/06}

\begin{document}


\title[Requirements for a Digital Library System: A Case Study in Digital Humanities (Technical Report)]{Requirements for a Digital Library System:\\ A Case Study in Digital Humanities (Technical Report)}


\author{Hermann Kroll}
\email{krollh@acm.org}
\orcid{0000-0001-9887-9276}
\affiliation{%
  \institution{Institute for Information Systems, TU Braunschweig}
  \city{Braunschweig}
  \country{Germany}
} 

\author{Christin K. Kreutz}
\email{ckreutz@acm.org}
\orcid{0000-0002-5075-7699}
\affiliation{%
  \institution{TH Mittelhessen \& Herder Institute}
  \city{Gießen \& Marburg}
\country{Germany}
}

\author{Mathias Jehn}
\author{Thomas Risse}
\email{{m.jehn, t.risse}@ub.uni-frankfurt.de}
\orcid{0000-0001-6248-1709}
\affiliation{%
  \institution{Goethe University Frankfurt, University Library}
  \city{Frankfurt}
  \country{Germany}
}

\renewcommand{\shortauthors}{Kroll et al.}

\begin{abstract}
Archives of libraries contain many materials, which have not yet been made available to the public. The prioritization of which content to provide and especially how to design effective access paths depend on potential users' needs. As a case study we interviewed researchers working on topics related to one German philosopher to map out their information interaction workflow. Additionally, we deeply analyze study participants' requirements for a digital library system. Moreover, we discuss how existing methods may meet their requirements and which implications these methods may have in a practical digital library setting, e.g., computational costs and hallucinations. In brief, this paper contributes the findings of our digital humanities case study resulting in system requirements.
\end{abstract}



\maketitle

\section{Introduction}
Archives of physical libraries are full of potentially interesting materials which have not yet been digitized~\cite{Ogilvie}. The process of digitization is still laborious, therefore there needs to be a prioritization of what to make available to the public~\cite{doi:10.1177/034003529802400205}. 
Digitization alone is not sufficient in enabling researchers to effectively access or explore the material - information access paths fitting the research interests or requirements should be constructed. 
In close cooperation with the University Library J. C. Senckenberg in Frankfurt am Main, we wanted to understand our users' requirements  when implementing access paths to one of our library's materials, e.g., to two collections of the famous philosophers Horkheimer and Schopenhauer. 
While digitizing the content and processing it with object character recognition (OCR) is the first step of making the content available, the question arose about what kind of access paths could help users in their daily research lives. Briefly, is it enough to provide keyword-based access? Or are more advanced access paths desired by our library's users? And if so, which requirements need to be met?

We tackled these questions in a case study. More precisely, we interviewed seven researchers from the sociological and philosophical domain who all worked on topics related to Schopenhauer. Our goal was to understand our users' information interaction process, so we asked them what kind of access paths they are using, how they search, and how they work with information. 

We tackle the research question \textit{(How) Can we support researchers from the philosophical and sociological domain?}
For this we make the following contributions: (1) We summarize our study setup so that other libraries can perform similar studies before implementing new access paths. (2) We summarize the key findings of our user study. We focus on shared workflows, used search paradigms and aspects that worked well or could be improved in researchers' current workflows. This summary can be seen as a case study within the digital humanities. (3) We discuss requirements for a digital library system when implementing access paths. Here, we divide requirements into two categories: (i) Requirements that can be met with existing tools/methods. (ii) Requirements like content translation, where language models show superior performance~\cite{DBLP:conf/eamt/MoslemHKW23,DBLP:journals/corr/abs-2401-06468}, but solutions are not yet ready~\cite{DBLP:journals/csur/JiLFYSXIBMF23} or straightforward to implement in a digital library.

\section{Related Work: Information Access Systems}

\subsubsection*{General Domain}
A multiplicity of traditional keyword-based digital library (DL) systems operate on a general domain such as the ACM DL, Bibsonomy~\cite{DBLP:series/xmedia/HothoJBGKSS09}, Google Scholar, Google Books, HathiTrust~\cite{DBLP:conf/chr/WalshLJCODD23} or 
Springer Link. 
Other systems based in Germany and directed towards the humanities are the German National Library's catalog OPAC\footnote{\url{https://portal.dnb.de/opac.htm}}, Karlsruhe Virtual Catalog (KVK)\footnote{\url{https://kvk.bibliothek.kit.edu/index.html?lang=en}} which is a meta search engine on 84 catalogs, 
Projekt Gutenberg-DE\footnote{\url{https://www.projekt-gutenberg.org}} containing historic German full texts and
Arcinsys Hessen\footnote{\url{https://arcinsys.hessen.de}} which is an archive information system in which materials can be ordered.
More sophisticated information access such as semantic search is possible with
Semantic Scholar~\cite{DBLP:conf/naacl/AmmarGBBCDDEFHK18}, 
Scopus AI\footnote{\url{https://www.elsevier.com/products/scopus/scopus-ai}, 05/2024} which additionally provides AI assistant features and
Connected Papers\footnote{\url{https://www.connectedpapers.com/}, 05/2024}.
Another access point are Academic Knowledge Graphs such as the ORKG~\cite{DBLP:journals/jodl/BrackHSAE22}, OpenAlex~\cite{DBLP:journals/corr/abs-2205-01833} or the already discontinued Microsoft Academic KG~\cite{DBLP:conf/www/SinhaSSMEHW15}. They allow sophisticated access through query languages like SPARQL which users must learn.

\subsubsection*{Domain-specific}
PhilPapers\footnote{\url{https://philpapers.org}} and Philosopher’s Index\footnote{\url{https://philindex.org}} index literature from the philosophical domain categorized into broad topics.
The Schopenhauer yearbooks~\cite{schopenhauer1912schopenhauer,koßler2023103} are a yearly appearing collection of research on the philosopher. The Thesaurus Schopenhauerianus\footnote{\url{https://www.schopenhauer-thesaurus.de}} indexes the content of yearbooks which are older than 5 years.
The Schopenhauer Archive provides original documents of Schopenhauer's family members, it can be accessed via Arcinsys.

\section{User Study: Interviews with In-Domain Researchers}
We investigate our research question \textit{(How) Can we support researchers from the philosophical and sociological domain?} through a small scale user study with domain experts to delve into the particularities of their information interactions. 

\subsection{Study Setup}
We invited seven senior researchers working on topics related to Schopenhauer to participate in our study.
For our study we scheduled 30 minute online sessions with each individual participant, an experienced interviewer and an observer taking notes. The sessions consisted of three parts: 1) introduction and consent ($\sim$ 5 minutes), 2) description of workflow ($\sim$ 15 minutes), 3) semi-structured interview ($\sim$ 10 minutes). The sessions were conducted in German during which the interviewer screen-shared slides holding the consent form and study questions.

Some days before the interviews, the participants received an email with the session setup, guide questions and the consent form. In part 1 of the study the interviewer and observer introduced themselves, the consent form was explained and the participant had the chance to ask initial questions. After verbally agreeing to the terms of the study, part 2 asked participants to first describe their current research questions before they then described their workflow how they \textit{tackle} these questions, i.e., how they search for and work with 
information related to their research question. Part 3 contained the following questions:
\begin{enumerate*}[label=(\roman*)]
    \item What do you think about your workflow: What is important to you? What works well? What is not yet working so well? 
    \item What would you wish for? How would you work under ideal conditions (in an ideal world)?
    \item What research questions would you like to answer that are difficult to address right now?
    \item Is there anything else you would like to add or ask?
\end{enumerate*}

\subsection{Information Interaction Workflows}
We consistently found four general components in the information interaction workflow throughout the diverse research questions the study participants work on, which will be described in more detail: 1) researchers use a multiplicity of \textit{information sources} such as platforms or libraries when 2) \textit{searching} for literature. They 3) assess \textit{relevancy} of the results they 4) finally \textit{work} with.

\paragraph{1. Information Sources.}
Researchers usually rely on multiple information sources. They use physical libraries and archives as well as newspapers and if the research question requires it even social media. For digital material, general purpose digital libraries such as Google Scholar, Google Books or HathiTrust were mentioned. More focused on catalogs we encountered mentions of the OPAC and the KVK. Other named general-purpose resources were Projekt Gutenberg-DE, the federal archive Germany, Arcinsys Hessen, and several university or state libraries.
Philosophy- or Schopenhauer-specific sources we encountered in the interviews were PhilPapers, the Philosopher's Index, the Thesaurus Schopenhauerianus, the Schopenhauer archive and Schopenhauer yearbooks.

\paragraph{2. Searching.}
When using the different information sources, strategies to obtain material range from simple keyword-based search over refinement of these keywords (sometimes by scanning results in different information sources) to combining keywords with the help of Boolean operators. Some researchers utilize the linkage of search results to other archives to find more results or check the actual content of relevant texts for links to other documents. Filtering of search results was regarded as helpful for some while it hindered others, e.g., a filter by publication year removing literature on a specific time period which appeared later. The same was observed for a semantic search– while it might help some to find related results not quite fitting the keywords, others detested the diluting of result lists for very focused queries where semantic search might fail. Hand-crafted topical ordering of literature as found in physical libraries was a feature positively mentioned by one study participant, as this enabled serendipity finds.

\paragraph{3. Result Lists and Relevancy.}
When researchers face result lists of potentially relevant literature, some go through each and every single result, others look at many results and a group of people only looks at few results deeply. There are study participants who even take multiple turns of going through result lists.
When investigating results, researchers mentioned a multiplicity of characteristics they might base their relevancy decisions on asides from its content: headlines or titles, keywords and/or semantics, publication date, the estate in which a material is stored, the content of a material’s detail page, the document type, author, publication source and reviews from third parties. It was also mentioned that observing material which one would consider irrelevant helps sharpen and reaffirms one’s notion of relevancy.

\paragraph{4. Working with Materials.}
When working with materials, there are three strategies: either having a thought process ready before starting to search for literature, doing the literature search before coming up with a mental concept or refining the initial concept while working though literature.
Literature is borrowed or archives and libraries are visited physically to read full texts. We heard about manually taking notes as well as annotating PDFs digitally. Typically, not all information is in the same language, so, an explicit translation step (e.g., using ChatGPT to translate an abstract from Latin to German to check if the relevancy estimation after a quick read of the original Latin text is correct) could take place.
With the information gained from the materials, one could directly write up an article, check biographic information on persons, manually classify the texts and make mind or manual maps of links between information.

\paragraph{Liked, Disliked and Desired Components.}
Researchers liked their own workflows. They praised the work of digital archives such as thesauri, the digitization of material, the availability of full texts and the enrichment of material with keywords.
Additionally, the possibility of using large language models (LLMs) for checking translations was liked.

As for disliked components, library-related aspects were mentioned such as the missing linkage of archives and meta search engines, required self-scans in physical libraries, and the non-straightforward process of finding the correct archive for a material.
Further, users disapproved of the inability to find relevant literature in different languages and searching for images in historical texts.

Desired components surpassing the improvement of current shortcomings were library-related technical ones such as voicing digitization requests online, obtaining citation information directly online and a bookmarking system with time stamps when a material was added a to list. Further, researchers wished for semantic search, an overview of literature on popular topics, the option to search for complex interactions between actors, layperson versions of natural science works and a topical filtering or grouping of material. A system with the option to link extracted information from materials was also requested.

\section{Requirements for a Digital Library System}
With the help of our interviews, we finally collected a list of requirements, that a future digital library system should fulfill.
We structure the requirements into following categories: 1. Technical requirements which can be solved by established methods today, 2. requirements which are tackled in research, for which ready-to-use solutions, that work beyond \textit{small demonstrations}, are yet missing. 

\subsection{Technical Requirements}

\paragraph{Digitizing the Content.}
First, the required content must be digitized, processed with suitable OCR tools like tesseract\footnote{\url{https://github.com/tesseract-ocr/tesseract}} or OCR-D, and stored so that users can access it through retrieval methods. We observed that keyword-based filtering would be a desirable goal. Often, users have to search online interfaces for possible items, go into the archives, and then check the material by hand. Here, a complete digitization of the content would be desirable.
Digitizing pure content is the first step, but the content needs to be enriched with metadata. Users wanted to filter the content by metadata like time, type, venue, author, and estate where material is stored.

\paragraph{Keyword-based Search.} Keyword-based search is a widely accepted paradigm for our users; see Apache Solr/Lucene\footnote{\url{https://solr.apache.org/}} for a practical implementation. It is easy to use and filters collections. When browsing and reading through material, keyword queries can be refined and improved. 

\paragraph{Federated Search.} 
We saw that our users worked with many different systems, mainly because the content was distributed. This process can be exhausting, especially when users access archives because they need to know where a certain piece is stored. At best, a digital library system should contain all of the related content. However, this is obviously a nearly impossible goal. That is why we argue that federated search or linking of information between platforms is the way to go here. Related information, for instance, to Schopenhauer, can then be provided if other systems are either automatically queried or at least links to related systems/searches are shown in the overarching federated system.

\subsection{Requirements and Possible Solutions}

\paragraph{Multilingual Retrieval.}
Research in the area of Schopenhauer is published mainly in German, English, French, and Italian. Some of our users stated that they read articles in different languages but that the search is challenging because each query must be formulated for every language (and in different systems). The requirement here is thus a cross-lingual retrieval component. For instance, the Europeana, Europe's largest platform for cultural heritage objects, faced a similar challenge, i.e., integrating multilingual content into one platform~\cite{DBLP:conf/ercimdl/MarreroI21}. The Europeana proposed to translate the query on-the-fly to retrieve content from different languages. The content is then shown in its original language. We argue that cross-lingual retrieval can be implemented today for \textit{similar} languages; see an implementation in the Europeana~\cite{DBLP:conf/ercimdl/MarreroI21} or in the medical domain~\cite{DBLP:conf/ecir/SalehP19}, and possible methods for cross-lingual retrieval~\cite{DBLP:conf/ecir/BalikasLRA18}, and for multilingual retrieval~\cite{DBLP:conf/ecir/LawrieYOM23}.

\paragraph{Language Translation.}
Our users stated that they work with content in different languages. For instance, one worked with old dissertations written in Latin. The person reads the dissertation in Latin briefly to get an impression of what is contained and then uses ChatGPT to translate those dissertations into German. If the content appears very relevant, the person translates the dissertation again by hand to ensure the quality of the translation. This example clearly shows the advantages and disadvantages of modern machine translation with large language models. On the one hand, translation helps to quickly grasp whether some content could be relevant to an information need. If so, the content can be translated by hand again to ensure the quality. On the other hand, translation must be of a suitable quality. The person argued that Latin-to-German translation would be fine because the \textit{person could read} Latin texts. The person would, however, not use such a translation for Arabic as the \textit{person}'s Arabic is not \textit{good enough}. 

We argue that with the advent neural machine translation and large language models, the quality of machine translation increased clearly and is \textit{nearly} ready to be used for library content. However, the limitations of such translations must be clearly shown: When using translation to get an impression of what is told, systems for high-resourced languages can be implemented today; see~\cite{DBLP:conf/eamt/MoslemHKW23,DBLP:journals/corr/abs-2401-06468} for evaluations of LLMs, \cite{DBLP:journals/csur/MarufSH21} for a survey of neural machine translation, or platforms like DeepL\footnote{\url{www.deepl.com}}. When expecting that no hallucinations or incorrect translations are generated, and thus, that translations do not need to be rechecked, technology is not ready yet. 
And beyond that, handling low-resourced languages with NLP methods remains challenging~\cite{DBLP:journals/coling/HaddowBBHB22,DBLP:conf/naacl/HedderichLASK21,DBLP:journals/csur/RanathungaLSSAK23}.

\paragraph{Content Exploration.}
We observed that users wanted to explore a library's content, e.g., start by searching for \textit{Schopenhauer AND Religion}, read some hits, and refine their search with terms related to or possibly replacing \textit{Religion}. In brief, users manually derived \textit{associated} terms to refine their searches. Existing approaches like finding associated terms through co-occurrences, word2vec~\cite{DBLP:journals/corr/abs-1301-3781} or generating keyword clouds with tools like YAKE~\cite{DBLP:journals/isci/CamposMPJNJ20,DBLP:conf/ecir/0001MPJNJ18a} could assist users in this process by displaying strongly related terms in the form of keyword clouds or some kind of autocompletion/suggestions as soon as they start typing a query. This way, they can get a quick impression of the library's content. In brief, methods like word2vec and keyword cloud generation have been researched for a while, and ready-to-use implementations like YAKE~\cite{DBLP:journals/isci/CamposMPJNJ20} exist. 
However, implementing them still limits users to express their information needs as keywords. 

More advanced access paths like question answering systems, e.g., Scopus AI, ChatGPT, and CORE-GPT~\cite{DBLP:conf/ercimdl/PrideCK23}, are becoming more present today. The main advantage from a user's perspective is that they can formulate their information needs as natural language questions. A system then replies with an answer formulated in natural language, sometimes accompanied by references/sources, e.g., see CORE-GPT. Users can, in this way, explore a library's content and refine their information needs by chatting with LLMs. However, we currently would still distinguish existing systems by their design: For instance, CORE-GPT~\cite{DBLP:conf/ercimdl/PrideCK23} takes a question, translates it into a keyword-based query, performs a search, and then replies with a list of possible related references. Users still have to work with the actual library content so that hallucinations are excluded by design. 
In contrast, LLMs like ChatGPT \textit{just} generate an answer that could be based on a spectrum from something real to purely hallucinated. Distinguishing what is real and what is not is still the question of research~\cite{DBLP:journals/csur/JiLFYSXIBMF23}. In brief, we argue that LLMs could help in transforming natural language questions into internal query representations, e.g., keyword queries, SQL, or SPARQL, but the result of the process should, at this moment, still be content of the library and not a generated answer that might contain hallucinations. Still, massive training data and computational costs are required to train and to apply a foundational models like LLaMa~\cite{DBLP:journals/corr/abs-2302-13971} or the open-source and open-data OLMo~\cite{DBLP:journals/corr/abs-2402-00838}. Additionally, related tasks like document summarization are also still facing the problem of hallucinations~\cite{DBLP:journals/csur/JiLFYSXIBMF23,DBLP:conf/ercimdl/KrollKCTB23,zhang-etal-2023-enhancing}. Even if a model comes with 99.9\% factual consistency regarding the library's content, would the quality be sufficient in practice? One wrongly generated answer could be one too much. As of today, we would argue methods like CORE-GPT~\cite{DBLP:conf/ercimdl/PrideCK23} are the best way for libraries to apply LLMs by \textit{still} retrieving the actual library content.
Another option related to content exploration desired by a study participant would be the creation of layperson versions of works from natural sciences. This could be achieved by the wide variety of text simplification approaches~\cite{DBLP:conf/clef/0002HKNS23,DBLP:conf/clef/ErmakovaBMK23,DBLP:conf/clef/WuH23a}.

\paragraph{Content Arrangement and Structuring.}
Content in a physical library is usually arranged by domain experts so that books are grouped based on similar topics. One person liked this way of exploring what else could be related to a specific article/book, e.g., by browsing through book/article titles with different terms or terminologies next to the initially found one in a physical library. Corresponding research and work for the digital world tries to mimic a physical library: Meghini et al.~\cite{DBLP:journals/amcs/MeghiniBMB19} demonstrated how narratives could be used to arrange content in the Europeana. More precisely, objects are arranged in a temporal and logical order so that a story about the corresponding artist is told. Sharaf et al. implemented a prototype to arrange papers~\cite{DBLP:conf/kdd/ShahafGH12} or news articles~\cite{DBLP:conf/kdd/ShahafG10} by \textit{connecting the dots}, i.e., they arranged articles on metro lines. A metro station represents an article, and a metro line represents a topic that connects different articles. Recently, we proposed a method to explore a Dutch library's newspaper archive by grouping articles by person aspects~\cite{DBLP:conf/ercimdl/KrollKCTB23}, e.g., what is told about the political career.

While those works demonstrate exciting opportunities for modern libraries to perform some kind of storytelling for users, the listed solutions should be seen as what they are: demonstrations. They focus on small examples and particular topics. Machine learning or manual curation is necessary to generate the story. That is why we argue solutions working across domains are still far-fetched.

\subsection{Discussion}
\textit{Limitations.}
We might have introduced a selection bias of interview partners because we only invited experienced researchers who themselves have experienced the difference of digitization on their research workflow.

\textit{Sustainable System Engineering.}
As we already argued, some requirements could be met by existing technologies. However, implementing sustainable systems is still challenging. First, digital libraries typically need to apply methods, that might have been proposed for small data sets, to their real-world collection. Especially the application of language models comes with high computation costs; see this study about costs of information retrieval methods~\cite{DBLP:conf/sigir/ScellsZZ22}. Second, digital libraries usually update their content, i.e., applied components must also be updated on a regular basis. While this sounds straightforward, the actual implications can be pretty expensive: Should we re-train some language models because one document has been added? Approaches like CORE-GPT~\cite{DBLP:conf/ercimdl/PrideCK23} sound promising: They combine well-established retrieval methods with a novel language model-driven question2keyword translation so that a re-training is not required for each content update. Our goal for the next step is thus to design a sustainable system through trade-offs: While the best quality might be achieved with some modern method, relying on an older one with less computational costs and better maintainability could be a better choice in practice.

\section{Conclusion}
In this work, we summarized the key findings of our user study in digital humanities. We derived a list of requirements that need to be met when implementing a digital library system. Moreover, we discussed in detail how some requirements, but not all, can be met with existing methods today and which consequences these methods may have in practice (e.g., computational costs, maintainability, and hallucinations). While we have yet to conclude a possible system design, this paper sheds light on requirements and solutions in digital humanities. In the future, we will continue our work by implementing a system for our users that should, at best, meet most requirements but is engineered sustainably.

\bibliographystyle{ACM-Reference-Format}
\bibliography{bib}

\end{document}